\documentclass[letterpaper, 10 pt, conference]{ieeeconf}  % Comment this line out if you need a4paper

\IEEEoverridecommandlockouts                 

\usepackage{cite}
\usepackage{amsmath,amssymb,amsfonts}
\usepackage{algorithmic}
\usepackage{graphicx}
\usepackage{textcomp}

\newtheorem{theorem}{\it Theorem}

\newtheorem{definition}{\it Definition}

\newtheorem{proposition}{\it Proposition}

\title{\LARGE \bf
Relativistic Rocket Control (Relativistic Space-Travel Flight Control): Feedback Control of Relativistic Dynamics Propelled by Ejecting Mass
}

\author{Song Fang and Quanyan Zhu% <-this % stops a space
\thanks{
Song Fang and Quanyan Zhu are with the Department of Electrical and Computer Engineering, New York University, New York, USA
{\tt \small song.fang@nyu.edu, quanyan.zhu@nyu.edu}%
}}

\begin{document}

\maketitle
\thispagestyle{empty}
\pagestyle{empty}

%%%%%%%%%%%%%%%%%%%%%%%%%%%%%%%%%%%%%%%%%%%%%%%%%%%%%%%%%%%%%%%%%%%%%%%%%%%%%%%%
\begin{abstract}

In this short note, we investigate the feedback control of relativistic dynamics propelled by mass ejection, modeling, e.g., the relativistic rocket control or the relativistic (space-travel) flight control. As an extreme case, we also examine the control of relativistic photon rockets which are propelled by ejecting photons.

\end{abstract}

%%%%%%%%%%%%%%%%%%%%%%%%%%%%%%%%%%%%%%%%%%%%%%%%%%%%%%%%%%%%%%%%%%%%%%%%%%%%%%%%
\section{Introduction}

In this short note, we investigate the feedback control of relativistic dynamics propelled by mass ejection, modeling, e.g., the feedback control of the relativistic rocket (or the relativistic flight in space travel; see, e.g.,  \cite{tinder2006relativistic, christodoulides2016special} and the references therein). 

We first introduce the state-space representation and  feedback linearization of classical dynamical systems propelled by mass ejection, based on the classical rocket equation \cite{tinder2006relativistic} that is derived from Newton's laws of motion \cite{feynman} and without considering the relativity effect. 

We then consider relativistic dynamics propelled by mass ejection, which obeys the relativistic rocket equation \cite{tinder2006relativistic, christodoulides2016special} that
takes into account of the special relativity \cite{feynman, einstein2013relativity, tolman1917theory, rindler1977essential, rindler2006relativity}. Based upon this, we examine the state-space representation and feedback linearization of the relativistic rocket. In addition, we investigate the feedback control of the relativistic rocket, providing discussions on its relativistic controllability, state-feedback control, output-feedback control, PID control, and so on. Finally, we study the case of relativistic photon rockets \cite{tinder2006relativistic, christodoulides2016special} for which the rockets are propelled by ejecting photons.

Note that parallel results have been presented in \cite{fang2019relativistic} on the feedback control of relativistic dynamics propelled by an external force, whereas the mass of the dynamical system does not vary over time. 

\section{Preliminaries}

\subsection{The Classical Rocket Equation and Its State-Space Representation}

Consider a rocket with mass $m \left( t \right) \in \mathbb{R} $. Denote its position as $p \left( t \right) \in \mathbb{R}$, its velocity as $ v \left( t \right) \in \mathbb{R}$, and its acceleration as $a \left( t \right) \in \mathbb{R}$. In addition, denote the velocity of the ejected mass (relative to the rocket) as $\overline{v} \in \mathbb{R}$, which is assumed to be constant and in the opposite direction of the rocket.
Then, according to the classical rocket equation \cite{tinder2006relativistic}, it holds that
\begin{flalign} \label{Newton}
\frac{ \mathrm{d}^2 p \left( t \right)}{\mathrm{d} t^2}  = \frac{ \mathrm{d} v \left( t \right)}{\mathrm{d} t}
= a \left( t \right) = -  \frac{ \overline{v} }{m \left( t \right)} \frac{ \mathrm{d} m \left( t \right)}{\mathrm{d} t}.
\end{flalign}
On the other hand, it is known \cite{tinder2006relativistic} that
\begin{flalign}
m \left( t \right) 
= m \left( 0 \right)  \mathrm{e}^{- \frac{v \left( t \right)}{\overline{v}}}.
\end{flalign}
In other words, \eqref{Newton} is equivalent to
\begin{flalign} \label{Newton2}
\frac{ \mathrm{d}^2 p \left( t \right)}{\mathrm{d} t^2}  = \frac{ \mathrm{d} v \left( t \right)}{\mathrm{d} t}
= a \left( t \right) = -  \mathrm{e}^{ \frac{v \left( t \right)}{\overline{v}}}  \frac{ \overline{v} }{m \left( 0 \right)} \frac{ \mathrm{d} m \left( t \right)}{\mathrm{d} t}.
\end{flalign}

In addition, by letting 
\begin{flalign}
\mathbf{x} \left( t \right) = 
\begin{bmatrix} 
p \left( t \right) \\
v \left( t \right) 
\end{bmatrix},~
u \left( t \right) = \frac{ \mathrm{d} m \left( t \right)}{\mathrm{d} t},
~
y \left( t \right) = p \left( t \right),
\end{flalign}
the following state-space representation of the system may be obtained:
\begin{flalign} 
\left\{ \begin{array}{rcl}
\frac{ \mathrm{d} \mathbf{x} \left( t \right)}{\mathrm{d} t} &=& f \left[  \mathbf{x} \left( t \right), u \left( t \right) \right], \\
y \left( t \right) &=& C \mathbf{x} \left( t \right),
\end{array}  
\right.
\end{flalign}
where
\begin{flalign}
f \left[  \mathbf{x} \left( t \right), u \left( t \right) \right]
=\begin{bmatrix} 
\begin{bmatrix} 
0 & 1
\end{bmatrix} \mathbf{x} \left( t \right) \\
-  \mathrm{e}^{ \frac{\left[ 	0 ~ 1 \right] \mathbf{x} \left( t \right)}{\overline{v}}}  \frac{ \overline{v} }{m \left( 0 \right)} u \left( t \right)
\end{bmatrix},
\end{flalign}
and
\begin{flalign}
C= \begin{bmatrix} 
1 & 0 
\end{bmatrix}.
\end{flalign}
This can be equivalently rewritten as
\begin{flalign} \label{state1}
\left\{ \begin{array}{rcl}
\dot{\mathbf{x}} \left( t \right) &=& A \mathbf{x} \left( t \right) + B \mathrm{e}^{ \frac{\left[ 	0 ~ 1 \right]
		\mathbf{x} \left( t \right)}{\overline{v}}} u \left( t \right), \\
y \left( t \right) &=& C \mathbf{x} \left( t \right),
\end{array}  
\right.
\end{flalign}
where 
\begin{flalign} \label{state2}
A = 
\begin{bmatrix} 
0 & 1 \\
0 & 0 
\end{bmatrix},
~B=\begin{bmatrix} 
0 \\
- \frac{\overline{v}}{m \left( 0 \right)} 
\end{bmatrix},
~C= \begin{bmatrix} 
1 & 0 
\end{bmatrix}.
\end{flalign}

We summarize the previous discussions of this subsection in the following proposition.

\begin{proposition}
	The state-space representation of the classical rocket equation \eqref{Newton} (or equivalently, \eqref{Newton2}) is given by \eqref{state1} and \eqref{state2}.
\end{proposition}

The system given in \eqref{state1} and \eqref{state2} can then be linearized \cite{isidori2013nonlinear, khalil2002nonlinear} with the following transformation:
\begin{flalign} \label{linearize11}
w \left( t \right) 
=  \mathrm{e}^{ \frac{\left[ 	0 ~ 1 \right]
		\mathbf{x} \left( t \right)}{\overline{v}}} u \left( t \right).
\end{flalign}
It can be verified that the transformation in \eqref{linearize11} is always invertible, and its inverse is given by
\begin{flalign} \label{linearize22}
u \left( t \right)
=  \mathrm{e}^{-  \frac{\left[ 	0 ~ 1 \right]
		\mathbf{x} \left( t \right)}{\overline{v}}} w \left( t \right).
\end{flalign}
In addition, both \eqref{linearize11} and \eqref{linearize22} are continuously differentiable. As such, the transformation in \eqref{linearize11} is a diffeomorphism and the system given in \eqref{state1} and \eqref{state2} is thus feedback linearizable \cite{isidori2013nonlinear, khalil2002nonlinear}.

In addition, with transformations \eqref{linearize11} and \eqref{linearize22}, the system from $w \left( t \right)$ to $y \left( t \right)$ will then be linear with its state-space model given by
\begin{flalign} \label{linearized11}
\left\{ \begin{array}{rcl}
\dot{\mathbf{x}} \left( t \right) &=& A \mathbf{x} \left( t \right) + B w \left( t \right), \\
y \left( t \right) &=& C \mathbf{x} \left( t \right),
\end{array}  
\right.
\end{flalign}
where
\begin{flalign} \label{linearized22}
A = 
\begin{bmatrix} 
0 & 1 \\
0 & 0 
\end{bmatrix},
~B=\begin{bmatrix} 
0 \\
- \frac{\overline{v}}{m \left( 0 \right)} 
\end{bmatrix},
~C= \begin{bmatrix} 
1 & 0 
\end{bmatrix}.
\end{flalign}

We now summarize the previous discussions on feedback linearization as follows.

\begin{proposition}
	If we choose $w \left( t \right)$ as in \eqref{linearize11} and \eqref{linearize22}, then the system given in \eqref{state1} and \eqref{state2} can be linearized as \eqref{linearized11} and \eqref{linearized22}.
\end{proposition}

As such, one may first design a controller for the linearized dynamics \cite{astrom}, and then this controller together with the feedback linearization compose the overall feedback controller design \cite{isidori2013nonlinear, khalil2002nonlinear}.

\subsection{The Relativistic Rocket Equation}

Consider a rocket with mass $m \left( \tau \right) \in \mathbb{R} $, where $\tau$ represents the time in the frame of reference of the rocket \cite{tinder2006relativistic, christodoulides2016special}. Denote its position as $p \left( t \right) \in \mathbb{R}$, its velocity as $ v \left( t \right) \in \mathbb{R}$, and its acceleration as $a \left( t \right) \in \mathbb{R}$, where $t$ represents the time in the frame of reference of the Earth \cite{tinder2006relativistic, christodoulides2016special}. Let $c$ denote the speed of light. In addition, denote the velocity of the ejected mass (relative to the rocket) as $\overline{v} \in \mathbb{R}$, measured in the frame of reference of the rocket \cite{tinder2006relativistic, christodoulides2016special}, which is assumed to be constant and in the opposite direction of the rocket.

It is known from \cite{tinder2006relativistic, christodoulides2016special} that in the Earth's frame of reference, the position $p \left( t \right)$, velocity $v \left( t \right)$, and acceleration $a \left( t \right)$ of the rocket are given by
\begin{flalign} \label{rrequation1}
\frac{\mathrm{d}^2 p \left( t \right) }{\mathrm{d} t^2}
&= \frac{\mathrm{d} v \left( t \right) }{\mathrm{d} t}
= a \left( t \right) \nonumber \\
&= - \frac{8 \overline{v}}{m \left( \tau \right) \left\{ \left[ \frac{m \left( \tau \right)}{m \left( 0 \right)}\right]^{\frac{\overline{v}}{c}} + \left[ \frac{m \left( \tau \right)}{m \left( 0 \right)} \right]^{\frac{-\overline{v}}{c}} \right\}^{3}} \frac{\mathrm{d} m \left( \tau \right)}{\mathrm{d} \tau},
\end{flalign}
which follows from the fact \cite{tinder2006relativistic, christodoulides2016special} that 
\begin{flalign}
\frac{v \left( t \right)}{c} &= \frac{1 - \left[ \frac{m \left( \tau \right)}{m \left( 0 \right)}\right]^{\frac{2\overline{v}}{c}}}{1 + \left[ \frac{m \left( \tau \right)}{m \left( 0 \right)}\right]^{\frac{2\overline{v}}{c}}}.
\end{flalign}
%where $\overline{v} = v \left( t \right) |_{t = 0}$, i.e., the initial velocity measured in the frame of reference of the Earth.
Note that herein the mass $m \left( \tau \right)$ and its derivative $\frac{\mathrm{d} m \left( \tau \right)}{\mathrm{d} \tau}$ are measured in the frame of reference of the rocket, in which the mass ejection is carried out.

On the other hand, the fact that
\begin{flalign}
\frac{v \left( t \right)}{c} &= \frac{1 - \left[ \frac{m \left( \tau \right)}{m \left( 0 \right)}\right]^{\frac{2\overline{v}}{c}}}{1 + \left[ \frac{m \left( \tau \right)}{m \left( 0 \right)}\right]^{\frac{2\overline{v}}{c}}}
\end{flalign}
is equivalent to \cite{tinder2006relativistic, christodoulides2016special}
\begin{flalign}
\frac{m \left( \tau \right)}{m \left( 0 \right)} = \left[ \frac{c - v \left( t \right)}{c + v \left( t \right)} \right]^{ \frac{c}{2\overline{v}}}. 
\end{flalign}
Accordingly, it holds that
\begin{flalign}
&\frac{8 \overline{v}}{m \left( \tau \right) \left\{ \left[ \frac{m \left( \tau \right)}{m \left( 0 \right)}\right]^{\frac{\overline{v}}{c}} + \left[ \frac{m \left( \tau \right)}{m \left( 0 \right)} \right]^{\frac{-\overline{v}}{c}} \right\}^{3} } \nonumber \\
&~~~~ = \frac{\overline{v}}{m \left( 0 \right) \left[ \frac{c - v \left( t \right)}{c + v \left( t \right)} \right]^{ \frac{c}{2\overline{v}}} \left[ 1 - \frac{v^2 \left( t \right)}{c^2} \right]^{-\frac{3}{2}}},
\end{flalign}
and thus 
\begin{flalign}  \label{rrequation2}
\frac{\mathrm{d}^2 p \left( t \right) }{\mathrm{d} t^2}
&= \frac{\mathrm{d} v \left( t \right) }{\mathrm{d} t}
= a \left( t \right) \nonumber \\
&= - \frac{\overline{v}}{m \left( 0 \right) \left[ \frac{c - v \left( t \right)}{c + v \left( t \right)} \right]^{ \frac{c}{2\overline{v}}} \left[ 1 - \frac{v^2 \left( t \right)}{c^2} \right]^{-\frac{3}{2}}} \frac{\mathrm{d} m \left( \tau \right)}{\mathrm{d} \tau}.
\end{flalign}

We summarize the previous discussions of this subsection in the following proposition.

\begin{proposition}
	The relativistic rocket equation given in \eqref{rrequation1} is equivalent to \eqref{rrequation2}.
\end{proposition}

Note that when $v \left( t \right) \ll c$, it can be verified that
\begin{flalign}
\left[ \frac{c - v \left( t \right)}{c + v \left( t \right)} \right]^{ \frac{c}{2\overline{v}}}
&= \left[ \frac{1 - \frac{v \left( t \right)}{c} }{1 + \frac{v \left( t \right)}{c} } \right]^{ \frac{c}{2\overline{v}}}
\approx \left[ 1 - \frac{2 v \left( t \right)}{c}  \right]^{ \frac{c}{2\overline{v}}} \nonumber \\
&= \left[ 1 - \frac{2 v \left( t \right)}{c}  \right]^{  \frac{-c}{2 v \left( t \right)} \frac{- v \left( t \right)}{\overline{v}}}
\approx \mathrm{e}^{ - \frac{ v \left( t \right)}{\overline{v}}},
\end{flalign}
and
\begin{flalign}
\left[ 1 - \frac{v^2 \left( t \right)}{c^2} \right]^{-\frac{3}{2}}
\approx 1 + \frac{3}{2} \frac{v^2 \left( t \right)}{c^2}
\approx  1,
\end{flalign}
while \cite{tinder2006relativistic, christodoulides2016special}
\begin{flalign}
\frac{\mathrm{d} m \left( \tau \right)}{\mathrm{d} \tau} \approx \frac{\mathrm{d} m \left( t \right)}{\mathrm{d} t}.
\end{flalign}
Hence, \eqref{rrequation2} reduces to
\begin{flalign}
\frac{ \mathrm{d}^2 p \left( t \right)}{\mathrm{d} t^2}  = \frac{ \mathrm{d} v \left( t \right)}{\mathrm{d} t}
= a \left( t \right) = -  \mathrm{e}^{ \frac{v \left( t \right)}{\overline{v}}}  \frac{ \overline{v} }{m \left( 0 \right)} \frac{ \mathrm{d} m \left( t \right)}{\mathrm{d} t},
\end{flalign}
and coincides with \eqref{Newton2}.
%On the other hand, when $v \left( t \right) \to c$, it holds that
%\begin{flalign}
%&\left[ \frac{c - v \left( t \right)}{c + v \left( t \right)} \right]^{ \frac{c}{2\overline{v}}} 
%\left[ 1 - \frac{v^2 \left( t \right)}{c^2} \right]^{- \frac{3}{2}} \nonumber \\
%&~~~~ =
%\left[ \frac{1 - \frac{v \left( t \right)}{c}}{1 + \frac{v \left( t \right)}{c}} \right]^{ \frac{c}{2\overline{v}}} 
%\left[ 1 - \frac{v^2 \left( t \right)}{c^2} \right]^{- \frac{3}{2}} \nonumber \\
%&~~~~ \to
%\left[ \frac{1 - \frac{v \left( t \right)}{c}}{2} \right]^{ \frac{c}{2\overline{v}}} 
%\left[ 1 - \frac{v^2 \left( t \right)}{c^2} \right]^{- \frac{3}{2}}
%\to \infty,
%\end{flalign}  
%indicating that it will require ejecting infinite mass to accelerate when the velocity approaches the speed of light \cite{}. As such, the difference between \eqref{rrequation2} and \eqref{Newton2} will tend to infinity in this extreme case, noting that when $v \left( t \right) \to c$,  \eqref{Newton2} reduces to
%\begin{flalign}
%\frac{ \mathrm{d}^2 p \left( t \right)}{\mathrm{d} t^2}  = \frac{ \mathrm{d} v \left( t \right)}{\mathrm{d} t}
%= a \left( t \right) = -  \mathrm{e}^{ \frac{c}{\overline{v}}}  \frac{ \overline{v} }{m \left( 0 \right)} \frac{ \mathrm{d} m \left( t \right)}{\mathrm{d} t},
%\end{flalign}
%where there exist no infinite terms.

\section{Feedback Control of the Relativistic Rocket}

In this section, we present the state-space representation and feedback control laws of the relativistic rocket. 

\subsection{State-Space Representation of the Relativistic Rocket}

By letting
\begin{flalign}
\mathbf{x} \left( t \right) = 
\begin{bmatrix} 
p \left( t \right) \\
v \left( t \right) 
\end{bmatrix},~u \left( \tau \right) = \frac{\mathrm{d} m \left( \tau \right)}{\mathrm{d} \tau},~y \left( t \right) = p \left( t \right),
\end{flalign}
we obtain the following state-space representation of \eqref{rrequation1} (or equivalently, \eqref{rrequation2}):
\begin{flalign} \label{ss1}
\left\{ \begin{array}{rcl}
\frac{\mathrm{d} \mathbf{x} \left( t \right)}{\mathrm{d} t} &=& f \left[  \mathbf{x} \left( t \right), u \left( \tau \right) \right], \\
y \left( t \right) &=& C \mathbf{x} \left( t \right),
\end{array}  
\right.
\end{flalign}
where
\begin{flalign}
&f \left[  \mathbf{x} \left( t \right), u \left( \tau \right) \right] \nonumber \\
&~~~~ =\begin{bmatrix} 
\begin{bmatrix} 
0 & 1 
\end{bmatrix} \mathbf{x} \left( t \right) \\
- \frac{\overline{v}}{m \left( 0 \right) \left[ \frac{c - \left[ 0~1 \right] \mathbf{x} \left( t \right)}{c + \left[ 0~1 \right] \mathbf{x} \left( t \right)} \right]^{ \frac{c}{2\overline{v}}} \left\{ 1 - \frac{ \left[ \left[ 0~1 \right] \mathbf{x} \left( t \right) \right]^2}{c^2} \right\}^{- \frac{3}{2}}} u \left( \tau \right)
\end{bmatrix} 
%\nonumber \\
%&~~~~ =\begin{bmatrix} 
%\left[ 0~1 \right] \mathbf{x} \left( t \right) \\
%- \left[ \frac{c - \left[ 0~1 \right] \mathbf{x} \left( t \right)}{c + \left[ 0~1 \right] \mathbf{x} \left( t \right)} \right]^{ - \frac{c}{2\overline{v}}} \left[ 1 - \frac{ \left[ \left[ 0~1 \right] \mathbf{x} \left( t \right) \right]^2}{c^2} \right]^{ \frac{3}{2}} \frac{\overline{v}}{m \left( 0 \right) } u \left( \tau \right)
%\end{bmatrix}
,
\end{flalign}
and
\begin{flalign}
C= \begin{bmatrix} 
1 & 0 
\end{bmatrix}.
\end{flalign}
Note that herein the control input $u \left( \tau \right)$ is defined in the frame of reference of the rocket \cite{tinder2006relativistic, christodoulides2016special}, in which the mass ejection is carried out. 
This representation can  equivalently be rewritten as
\begin{flalign} \label{Rstate1}
\left\{ \begin{array}{rcl}
\dot{\mathbf{x}} \left( t \right) &=& A \mathbf{x} \left( t \right) +  \frac{B u \left( \tau \right)}{ \left[ \frac{c - \left[ 0~1 \right] \mathbf{x} \left( t \right)}{c + \left[ 0~1 \right] \mathbf{x} \left( t \right)} \right]^{ \frac{c}{2\overline{v}}} \left\{ 1 - \frac{ \left[ \left[ 0~1 \right] \mathbf{x} \left( t \right) \right]^2}{c^2} \right\}^{- \frac{3}{2}}}, \\
y \left( t \right) &=& C \mathbf{x} \left( t \right),
\end{array}  
\right.
\end{flalign}
where
\begin{flalign} \label{Rstate2}
A = 
\begin{bmatrix} 
0 & 1 \\
0 & 0 
\end{bmatrix},
~B=\begin{bmatrix} 
0 \\
- \frac{\overline{v}}{m_{0}} 
\end{bmatrix},
~C= \begin{bmatrix} 
1 & 0 
\end{bmatrix},
\end{flalign}
and it may be viewed as the state-space representation of the relativistic rocket.

We summarize the previous discussions of this subsection in the following proposition.

\begin{proposition}
	The state-space representation of the relativistic rocket equation \eqref{rrequation1} (or equivalently, \eqref{rrequation2}) is given by \eqref{Rstate1} and \eqref{Rstate2}.
\end{proposition}

\subsection{Feedback Linearization of the Relativistic Rocket}

We will now linearize \cite{isidori2013nonlinear, khalil2002nonlinear} the system given in \eqref{Rstate1} and \eqref{Rstate2} with the following transformation:
\begin{flalign} \label{linearize1}
w \left( \tau \right) 
=  \frac{1}{ \left[ \frac{c - \left[ 0~1 \right] \mathbf{x} \left( t \right)}{c + \left[ 0~1 \right] \mathbf{x} \left( t \right)} \right]^{ \frac{c}{2\overline{v}}} \left\{ 1 - \frac{ \left[ \left[ 0~1 \right] \mathbf{x} \left( t \right) \right]^2}{c^2} \right\}^{- \frac{3}{2}}}  u \left( \tau \right).
\end{flalign}
It can be verified that the transformation in \eqref{linearize1} is invertible for $\left| v \left( t \right) \right| = \left| \begin{bmatrix} 
0 & 1
\end{bmatrix} \mathbf{x} \left( t \right) \right| < c$, and its inverse is given by
\begin{flalign} \label{linearize2}
u \left( \tau \right)
=  \left[ \frac{c - \left[ 0~1 \right] \mathbf{x} \left( t \right)}{c + \left[ 0~1 \right] \mathbf{x} \left( t \right)} \right]^{ \frac{c}{2\overline{v}}} \left\{ 1 - \frac{ \left[ \left[ 0~1 \right] \mathbf{x} \left( t \right) \right]^2}{c^2} \right\}^{- \frac{3}{2}} w \left( \tau \right).
\end{flalign}
In addition, both \eqref{linearize1} and \eqref{linearize2} are continuously differentiable for $\left| v \left( t \right) \right|  < c$. As such, the transformation in \eqref{linearize1} is a diffeomorphism and the system given in \eqref{Rstate1} and \eqref{Rstate2} is thus feedback linearizable \cite{isidori2013nonlinear, khalil2002nonlinear} for $\left| v \left( t \right) \right| < c$.

On the other hand, it is known \cite{tinder2006relativistic, christodoulides2016special} that $\left| v \left( t \right) \right| < c$ will always hold. In this sense, the system is always relativistically feedback linearizable, and we may always linearize the system using \eqref{linearize1} and \eqref{linearize2}.

%relativistically feedback linearizable (a notion?)

As a matter of fact, with transformations \eqref{linearize1} and \eqref{linearize2}, the system from $w \left( \tau \right)$ to $y \left( t \right)$ will then be linear with its state-space model given by
\begin{flalign} \label{linearized1}
\left\{ \begin{array}{rcl}
\dot{\mathbf{x}} \left( t \right) &=& A \mathbf{x} \left( t \right) + B w \left( \tau \right), \\
y \left( t \right) &=& C \mathbf{x} \left( t \right),
\end{array}  
\right.
\end{flalign}
where
\begin{flalign} \label{linearized2}
A = 
\begin{bmatrix} 
0 & 1 \\
0 & 0 
\end{bmatrix},
~B=\begin{bmatrix} 
0 \\
- \frac{\overline{v}}{m_{0}} 
\end{bmatrix},
~C= \begin{bmatrix} 
1 & 0 
\end{bmatrix}.
\end{flalign}

We now summarize the previous discussions of this subsection in the following theorem.

\begin{theorem}
	If we choose $w \left( \tau \right)$ as in \eqref{linearize1} and \eqref{linearize2}, then the system given in \eqref{Rstate1} and \eqref{Rstate2} can be linearized as \eqref{linearized1} and \eqref{linearized2}.
\end{theorem}

%As such, we may first design a controller for the linearized dynamics \eqref{linearized1} and \eqref{linearized2}. Then, this controller together with the feedback linearization compose the overall feedback controller design for the relativistic dynamics given in \eqref{Rstate1} and \eqref{Rstate2}. Detailed discussions on this as well as properties such as controllability will be presented in what follows.

\subsection{Relativistic Rocket Control} \label{control}

In this subsection, we investigate the feedback control of the relativistic rocket.

\subsubsection{Relativistic Controllability}

It can be verified that the system given in \eqref{state1} and \eqref{state2} is controllable, whereas the system given in \eqref{Rstate1} and \eqref{Rstate2} is not controllable \cite{astrom, dorf2011modern} since any state $\mathbf{x} \left( t \right)$ with $\left| v \left( t \right) \right| > c$ is not reachable. This is a fundamental difference between the classical rocket equation and the relativistic rocket equation.

It may then be verified that although the system given in \eqref{Rstate1} and \eqref{Rstate2} is not controllable, it is relativistically controllable \cite{fang2019relativistic}, i.e, any state $\mathbf{x} \left( t \right)$ with $\left| v \left( t \right) \right| < c$ is reachable. To see this, note first that the system given in \eqref{linearized1} and \eqref{linearized2} is controllable. Hence, the system input $w \left( \tau \right): \left[ t_0,~T \right] \to \mathbb{R}$ can be designed to steer any initial state $\mathbf{x} \left( t_0 \right)$ to any other final state $\mathbf{x} \left( T \right)$ in a finite time. As a special case, it will certainly be possible to design a $w \left( \tau \right): \left[ t_0,~T \right] \to \mathbb{R}$ to steer any initial state $\mathbf{x} \left( t_0 \right)$ with $\left| v  \left( t \right) \right| < c$ to any other final state $\mathbf{x} \left( T \right)$ with $\left| v  \left( t \right) \right| < c$ in a finite time. Correspondingly, with $u \left( \tau \right): \left[ t_0,~T \right] \to \mathbb{R}$, where
\begin{flalign} 
u \left( \tau \right)
=  \left[ \frac{c - v \left( t \right)}{c + v \left( t \right)} \right]^{ \frac{c}{2\overline{v}}} \left[ 1 - \frac{ v^2 \left( t \right)}{c^2} \right]^{- \frac{3}{2}} w \left( \tau \right),~\left| v  \left( t \right) \right| < c,
\end{flalign}
any initial state $\mathbf{x} \left( t_0 \right)$ with $\left| v  \left( t \right) \right| < c$ can thus be steered to any other final state $\mathbf{x} \left( T \right)$ with $\left| v  \left( t \right) \right| < c$ in a finite time for the system given by \eqref{Rstate1} and \eqref{Rstate2}.

\subsubsection{Relativistic Rocket State Feedback Control}

If the state feedback control law for the system given in \eqref{linearized1} and \eqref{linearized2} is designed as \cite{astrom}
\begin{flalign}
w \left( \tau \right)
= - K  \mathbf{x} \left( t \right),
\end{flalign}
then, noting also that
\begin{flalign}
u \left( \tau \right)
= \left[ \frac{c - \left[ 0~1 \right] \mathbf{x} \left( t \right)}{c + \left[ 0~1 \right] \mathbf{x} \left( t \right)} \right]^{ \frac{c}{2\overline{v}}} \left\{ 1 - \frac{ \left[ \left[ 0~1 \right] \mathbf{x} \left( t \right) \right]^2}{c^2} \right\}^{- \frac{3}{2}} w \left( \tau \right),
\end{flalign}
the overall relativistic state feedback controller is given by
\begin{flalign} \label{Rstatefeedback}
u \left( \tau \right)
&=  - \left[ \frac{c - \left[ 0~1 \right] \mathbf{x} \left( t \right)}{c + \left[ 0~1 \right] \mathbf{x} \left( t \right)} \right]^{ \frac{c}{2\overline{v}}} \nonumber \\
&~~~~ \times \left\{ 1 - \frac{ \left[ \left[ 0~1 \right] \mathbf{x} \left( t \right) \right]^2}{c^2} \right\}^{- \frac{3}{2}} K  \mathbf{x} \left( t \right).
\end{flalign}

We summarize the discussions on state feedback in the following definition.

\begin{definition}
	The state feedback controller for the relativistic rocket is given by \eqref{Rstatefeedback}.
\end{definition}

\subsubsection{Relativistic Rocket Output Feedback Control}

If the output feedback control law for the system given in \eqref{linearized1} and \eqref{linearized2} is designed as \cite{astrom}
\begin{flalign}
w \left( \tau \right)
= l \left[ y \left( t \right) \right],
\end{flalign}
then, noting as well that
\begin{flalign}
u \left( \tau \right)
= \left[ \frac{c - \left[ 0~1 \right] \mathbf{x} \left( t \right)}{c + \left[ 0~1 \right] \mathbf{x} \left( t \right)} \right]^{ \frac{c}{2\overline{v}}} \left\{ 1 - \frac{ \left[ \left[ 0~1 \right] \mathbf{x} \left( t \right) \right]^2}{c^2} \right\}^{- \frac{3}{2}} w \left( \tau \right),
\end{flalign}
with
\begin{flalign}
\begin{bmatrix} 
0 & 1
\end{bmatrix} \mathbf{x} \left( t \right)
= \frac{\mathrm{d} y \left( t \right)}{\mathrm{d} t},
\end{flalign}  
the overall relativistic output-feedback controller is given by
\begin{flalign} \label{output}
u \left( \tau \right)
=  \left[ \frac{c - \frac{\mathrm{d} y \left( t \right)}{\mathrm{d} t}}{c + \frac{\mathrm{d} y \left( t \right)}{\mathrm{d} t} } \right]^{ \frac{c}{2\overline{v}}} \left\{ 1 - \frac{ \left[ \frac{\mathrm{d} y \left( t \right)}{\mathrm{d} t} \right]^2}{c^2} \right\}^{- \frac{3}{2}} l \left[ y \left( t \right) \right].
\end{flalign}

We next summarize the discussions concerning output feedback in the following definition.

\begin{definition}
	The output feedback controller for the relativistic rocket is given by \eqref{output}.
\end{definition}

\subsubsection{Relativistic Rocket PID Control}

We now consider a special case of output feedback: PID control. Suppose that $w \left( \tau \right)$ is designed using PID control as \cite{astrom}
\begin{flalign}
w(t)=K_{\text{p}} e \left( t \right) + K_{\text{i}} \int_{0}^{t}e \left( \tau \right) \text{d}\tau+K_{\text{d}} \frac{\mathrm{d} e \left( t \right)}{\mathrm{d} t},
\end{flalign}
where 
\begin{flalign} 
e \left( t \right) = r - y \left( t \right).
\end{flalign} 
Note that herein $r$ is given in the frame of reference of the Earth \cite{tinder2006relativistic, christodoulides2016special}.
On the other hand, 
\begin{flalign}
\begin{bmatrix} 
0 & 1
\end{bmatrix} \mathbf{x} \left( t \right) = \frac{\mathrm{d} y \left( t \right)}{\mathrm{d} t} =  - \frac{\mathrm{d} e \left( t \right)}{\mathrm{d} t}.
\end{flalign}  
As such, noting also that
\begin{flalign}
u \left( \tau \right)
= \left[ \frac{c - \left[ 0~1 \right] \mathbf{x} \left( t \right)}{c + \left[ 0~1 \right] \mathbf{x} \left( t \right)} \right]^{ \frac{c}{2\overline{v}}} \left\{ 1 - \frac{ \left[ \left[ 0~1 \right] \mathbf{x} \left( t \right) \right]^2}{c^2} \right\}^{- \frac{3}{2}} w \left( \tau \right),
\end{flalign}
the overall relativistic PID controller is given by
\begin{flalign} \label{RPID}
u \left( \tau \right)  
& =  \left[ \frac{c - \frac{\mathrm{d} e \left( t \right)}{\mathrm{d} t}}{c + \frac{\mathrm{d} e \left( t \right)}{\mathrm{d} t}} \right]^{ \frac{c}{2\overline{v}}} \left\{ 1 - \frac{ \left[ \frac{\mathrm{d} e \left( t \right)}{\mathrm{d} t} \right]^2}{c^2} \right\}^{- \frac{3}{2}} \nonumber \\
&~~~~ \times \left[ K_{\text{p}} e \left( t \right) + K_{\text{i}} \int_{0}^{t}e \left( \tau \right) \text{d}\tau+K_{\text{d}} \frac{\mathrm{d} e \left( t \right)}{\mathrm{d} t} \right].
\end{flalign}

We summarize the discussions on PID control in the following definition.

\begin{definition}
	The PID controller for the relativistic rocket is given by \eqref{RPID}.
\end{definition}

Note that herein we assumed that $r$ is a constant and thus $ \frac{ \mathrm{d} r }{\mathrm{d} t} = 0$. If this is not the case, we can invoke the fact
\begin{flalign}
\begin{bmatrix} 
0 & 1
\end{bmatrix} \mathbf{x} \left( t \right) = \frac{ \mathrm{d} y \left( t \right) }{\mathrm{d} t}  = \frac{ \mathrm{d} r \left( t \right) }{\mathrm{d} t} - \frac{ \mathrm{d} e \left( t \right) }{\mathrm{d} t},
\end{flalign} 
and obtain a corresponding result.

\subsection{The Relativistic Photon Rocket}

For a photon rocket that is propelled by ejecting photons, it holds that $\overline{v} = c$ \cite{tinder2006relativistic, christodoulides2016special}. Accordingly, \eqref{rrequation2} reduces to
\begin{flalign}
\frac{\mathrm{d}^2 p \left( t \right) }{\mathrm{d} t^2}
&= \frac{\mathrm{d} v \left( t \right) }{\mathrm{d} t}
= a \left( t \right) \nonumber \\
&= - \frac{\overline{v}}{m \left( 0 \right) \left[ \frac{c - v \left( t \right)}{c + v \left( t \right)} \right]^{\frac{1}{2}} \left[ 1 - \frac{v^2 \left( t \right)}{c^2} \right]^{\frac{3}{2}}} \frac{\mathrm{d} m \left( \tau \right)}{\mathrm{d} \tau},
\end{flalign}
and \eqref{Rstate1} becomes
\begin{flalign} 
\left\{ \begin{array}{rcl}
\dot{\mathbf{x}} \left( t \right) &=& A \mathbf{x} \left( t \right) +  \frac{B u \left( \tau \right)}{ \left[ \frac{c - \left[ 0~1 \right] \mathbf{x} \left( t \right)}{c + \left[ 0~1 \right] \mathbf{x} \left( t \right)} \right]^{ \frac{1}{2}} \left\{ 1 - \frac{ \left[ \left[ 0~1 \right] \mathbf{x} \left( t \right) \right]^2}{c^2} \right\}^{- \frac{3}{2}}}, \\
y \left( t \right) &=& C \mathbf{x} \left( t \right).
\end{array}  
\right.
\end{flalign}
In addition, \eqref{linearize2} reduces to
\begin{flalign} 
u \left( \tau \right)
=  \left[ \frac{c - \left[ 0~1 \right] \mathbf{x} \left( t \right)}{c + \left[ 0~1 \right] \mathbf{x} \left( t \right)} \right]^{ \frac{1}{2}} \left\{ 1 - \frac{ \left[ \left[ 0~1 \right] \mathbf{x} \left( t \right) \right]^2}{c^2} \right\}^{- \frac{3}{2}} w \left( \tau \right).
\end{flalign}
Correspondingly, the feedback control methods as discussed in Section~\ref{control} are still applicable herein for the photon rocket, after letting $\overline{v} = c$.

\section{Conclusions}

In this note, we have introduced the state-space representation of the relativistic rocket and investigated how to design the relativistic rocket controller based on feedback linearization.  Potential future research directions include analysis and design of the relativistic rocket (state) estimator. 

\bibliographystyle{IEEEtran}
\bibliography{references}

\end{document}